\documentclass[11pt]{article}
\usepackage{fullpage, amsmath, amsfonts, amsthm, amssymb, natbib, tikz, bbm}
\usepackage{graphicx, enumerate, color, hyperref}
\usepackage{algorithm, algorithmic}
\usepackage{mathtools}
\usepackage{multirow}
\usepackage{float}
\usepackage{array}
\usepackage{caption}
\usepackage{hyperref}
\usepackage[flushleft]{threeparttable}
\usepackage{setspace}
\usepackage{tabularx}
\usepackage{bm}
\usepackage{subcaption}
\theoremstyle{definition}

\theoremstyle{remark}

\newcommand{\argmin}{\operatornamewithlimits{argmin}}
\newcommand{\utwi}[1]{\mbox{\boldmath $ #1$}}

\newcommand{\Z}{{\utwi{Z}}}
\newcommand{\PhiB}{{\utwi{\Phi}}}

\title{An Improved Online Penalty Parameter Selection Procedure for $\ell_1$-Penalized Autoregressive with Exogenous Variables}
\author{William B. Nicholson\thanks{Research Lead, Point 72 Asset Management, L.P., New York, NY (Email: \href{mailto:wbn8@cornell.edu}{wbn8@cornell.edu}; Webpage: \url{http://www.wbnicholson.com}).  Mr. Nicholson contributed to this article in his personal capacity. The information, views, and opinions expressed herein are solely his own and do not necessarily represent the views of Point72. Point72 is not responsible for, and did not verify for accuracy, any of the information contained herein.} \and Xiaohan Yan\thanks{Data Scientist, Microsoft Azure, Redmond, WA (Email: \href{mailto:xy257@cornell.edu}{xy257@cornell.edu}; Webpage: \url{https://yanxht.github.io}). Mr. Yan contributed to this article in his personal capacity. The information, views, and opinions expressed herein are solely his own and do not necessarily represent the views of Microsoft. Microsoft is not responsible for, and did not verify for accuracy, any of the information contained herein.}}
\date{\today}

\begin{document}
\maketitle
\begin{abstract}
Many recent developments in the high-dimensional statistical time series literature have centered around time-dependent applications that can be adapted to regularized least squares.  Of particular interest is the lasso, which both serves to regularize and provide feature selection.  The lasso requires the specification of a penalty parameter that determines the degree of sparsity to impose.  The most popular penalty parameter selection approaches that respect time dependence are very computationally intensive and are not appropriate for modeling certain classes of time series.  We propose enhancing a canonical time series model, the autoregressive model with exogenous variables, with a novel online penalty parameter selection procedure that takes advantage of the sequential nature of time series data to improve both computational performance and forecast accuracy relative to existing methods in both a simulation and empirical application involving macroeconomic indicators.      
\end{abstract}


\section{Introduction}
\label{sec:sec1}
Selecting relevant features in a time series model is a longstanding open problem in both statistics and econometrics.  Traditional feature selection approaches, such as those proposed by \cite{BoxJenkins} rely on heuristic methods, such as visual inspection of diagnostic plots.  Alternatively, popular data-driven approaches select features based upon the minimization of information criterion, such as Akaike's Information Criterion (AIC, \citealt{ak}) or Bayesian Information Criterion (BIC, \citealt{schwarz}) over a subset of potential models.  Typically, because the space of all possible models is extremely large, such methods require imposing substantial restrictions on the feature space.

More recent approaches have extended the \textit{lasso} \citep{tibs} to a time dependent setting.  The lasso has advantages over conventional methods in that it shrinks least squares estimates toward zero in addition to performing feature selection.  It also allows for estimation under scenarios in which there are more potential features than observations.  The lasso utilizes an $\ell_1$ penalty parameter to control the tradeoff between the least squares objective and the degree of regularization.  In most scenarios there is no theoretical basis for choosing this parameter, so it must be estimated empirically.  A standard selection approach, $n$-fold cross validation, does not respect time dependence.  ``Rolling''  validation, a popular approach used by \cite{Nicholson}, \cite{BickelSong}, \cite{BGR}, and \cite{Koop} involves incrementing the data forward one observation at a time over a training period and selecting the penalty parameter as the minimizer of a loss function (typically MSFE) over a grid of candidate penalty values.  This procedure is very computationally intensive, as it requires recalculating the lasso solution at every time point despite adding just one observation.  

Moreover, the ``optimal'' value reported from a fixed grid of penalty parameters may not be appropriate in modeling certain classes of time-dependent problems; in particular nonstationary series in which parameter relationships may vary across time.  Penalty parameter selection procedures designed to accommodate nonstationary time series should be able to adapt to varying degrees of dependence.       

Through the incorporation of an online procedure proposed by \cite{Garr} which updates the current lasso solution as a new observation is received rather than completely re-estimating, we are able to substantially improve upon the computational performance of the lasso.  While adapting this procedure to a time-dependent setting, we present an online updating scheme that allows the penalty parameter to dynamically adjust in the presence of new information in order to better accommodate a larger class of time dependent problems.  This substantially improves upon out-of-sample forecast performance for both simulated data and several macroeconomic forecasting applications.  

Section \ref{sec:sec2} introduces the autoregressive with exogenous variables (AR-X) framework that is used throughout the paper, explains the incorporation of the $\ell_1$ penalty and addresses the issues with the conventional penalty parameter selection procedure. Section \ref{sec:sec3} presents the online updating scheme used to improve computational performance and our proposed online regularization scheme.  Section \ref{sec:sec4} details our results in both simulations and a macroeconomic data application, and Section \ref{sec:sec5} contains our conclusion.   

\section{Methodology}
\label{sec:sec2}
In this section, we provide a brief overview of autoregressive modeling with exogenous variables.  We start by demonstrating that the $\ell_1$-penalized AR-X offers greater flexibility than conventional information criterion based methods in selecting relevant features.  We additionally provide an overview of rolling validation and detail some of its shortcomings.  

\subsection{The AR-X Framework}
Our analysis operate in the context of autoregressive processes with exogenous variables (AR-X), a canonical time series model that is amenable toward applying a lasso penalty, as it can be formulated as a least squares problem.  We consider forecasting the length $T$ series $\{y_t\}_{t=1}^T$ using its $p$ most recent lagged values $[y_{t-1},\ldots,y_{t-p}]$ as well as $s$ lagged values of $\{\bm{x}_t\}_{t=1}^T $ from $\mathbb{R}^{k}$, which represent $k$ unmodeled, exogenous series.  In the context of this paper, a time series is considered \emph{exogenous} if it is not modeled, but it aids in forecasting our series of interest $\{y_t\}_{t=1}^T$.  For example, if we are forecasting the US Federal Funds Rate, potential exogenous series can include other relevant macroeconomic indicators, such as the Consumer Price Index or Gross Domestic Product growth rate.   
  
Typically, maximal lag orders (i.e. the maximum number of lagged features included in the model) for both $y_t$ and $\bm{x}_t$ are chosen according to the frequency of the data.  For example if the series are recorded quarterly, one might choose $p=s=4$ to represent one year of past dependence; for monthly data, one might select $p=s=12$.  

An AR-X$(p,s)$ model at time $t$ can be estimated by least squares via the objective function 
\begin{equation}
\label{eqn1}
\min_{\bm{\phi}\in \mathbb R^p,\bm{\theta}\in \mathbb R^{ks}} \frac{1}{2} \sum_{i=T_0}^t\left(y_i-\sum_{j=1}^p\phi_{j}y_{i-j}-\sum_{\ell=1}^k\sum_{j=1}^s\theta_{\ell j}x_{\ell,i-j}\right)^2
\end{equation}
in which $\phi_j, \theta_{\ell j} \in\mathbb{R}$ denote model coefficients and $T_0=\max(p,s)$ offsets the values needed to create the AR-X lag matrix. 

\subsection{Feature Selection: from IC Methods to the Lasso} 
In general, it is not desirable to report forecasts from an AR-X$(p,s)$ if $p$ and $s$ are large relative to $T$ due to concerns of overfitting.  Since the least squares objective function decreases monotonically as the number of features included increases, it is natural to apply a penalty that restricts the feature space.  

Conventional penalized approaches involve the use of information criterion (henceforth IC).  Rooted in information theory, IC based methods, such as AIC and BIC provide a coherent framework for feature selection.  A standard technique involves fitting several nested models over a subset of the feature space and selecting the model that minimizes a chosen IC. As stated in \cite{Hsu}, the space of all potential models can be extremely large.  Even if $p,k$ and $s$ are relatively small, it is not computationally tractable to fit each of the possible $2^{p+ks}$ subset models.  

In the IC setting, the model space is typically restricted by assuming that all coefficients are nonzero up to maximal lag orders $\hat{p}$ and $\hat{s}$.  These lag orders are typically selected by fitting a least squares problem (e.g., Problem \ref{eqn1}) for an AR-X for $0\leq \tilde p \leq p$, $0\leq \tilde s \leq s$ and selecting the optimal $\tilde p$ and $\tilde {s}$ based on the minimization of an IC.  The AIC and BIC of an AR-X$(\tilde p, \tilde s)$ are defined as:
\begin{equation*}
\text{AIC}(\tilde p, \tilde s)=\log(\hat{\sigma}_{u}^{\tilde p, \tilde s})+\frac{2(\tilde p + k\tilde s)}{T} \quad\text{and}\quad\text{BIC}(\tilde p, \tilde s)=\log(\hat{\sigma}_{u}^{\tilde p, \tilde s})+\frac{\log(T)(\tilde p + k\tilde s)}{T} 
\end{equation*}
in which $\hat{\sigma}_{u}^{\tilde p, \tilde s}$ is the squared residual obtained from using Problem \ref{eqn1} to fit an AR-X$(\tilde p, \tilde s)$. As AIC penalizes model coefficients uniformly by a factor of 2 whereas BIC scales penalties according to series length, BIC tends to select more parsimonious models (i.e. smaller maximum lag orders) than AIC does.  

Since we are limited to a nested subset of potential models, IC based methods can be very restrictive.  In particular, it is typically assumed that every exogenous series in $\bm{x}_t$ has the same maximal lag order, although \cite{penm} shows that this rarely holds empirically. In contrast, the lasso solution can be obtained by adding the $\ell_1$ penalty   
\begin{equation}
  \lambda\left(\sum_{j=1}^p|\phi_j|+\sum_{\ell=1}^k\sum_{j=1}^{s}|\theta_{\ell j}|\right)
\end{equation}
to the objective function in Problem \eqref{eqn1}, resulting in a lasso AR-X problem.  We denote $\lambda\geq 0$ as the parameter that controls the tradeoff between the least squares fit and the degree of regularization.  As opposed to other penalized regression methods, such as ridge regression, due to the inclusion of the $\ell_1$ penalty, the lasso returns a sparse set of coefficients.  Larger values of $\lambda$ tend to lead to more sparse solutions.  In contrast to IC methods, the lasso performs estimation and feature selection in one step and does not require imposing any restrictions on the feature space. 

\begin{figure}
\centering
\includegraphics[scale=.66]{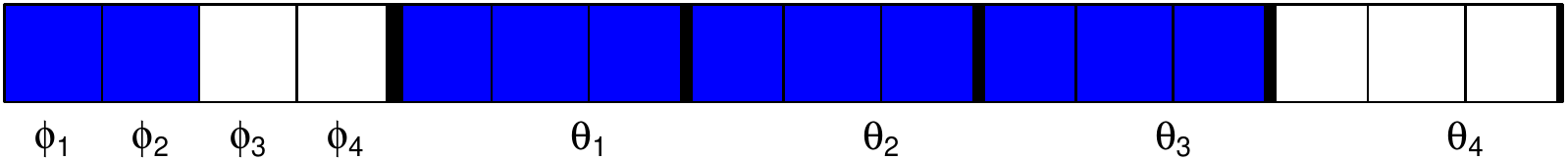}
\includegraphics[scale=.66]{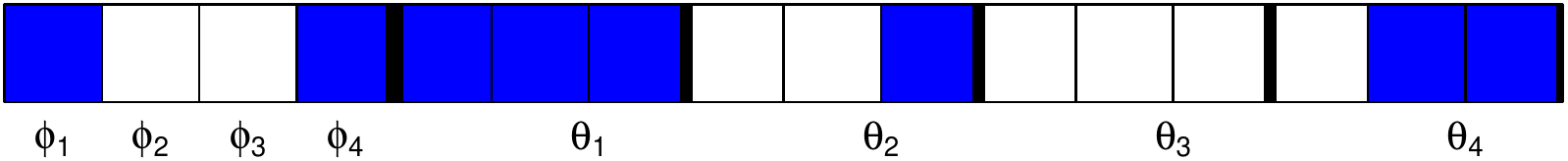}
\caption{Example features selected (shaded) from an AR-X(4,4) with 3 exogenous series.  The top figure corresponds to features selected by IC minimization whereas the bottom corresponds to the lasso. }
\label{fig:fig1}
\end{figure}

Figure \ref{fig:fig1} depicts an example highlighting the differences in features selected by IC minimization and the lasso.  The lasso can select any subset of features, hence it can more accurately capture the non-monotonic feature relationships that are common in time series, such as seasonal dependence (e.g. nonzero coefficients at every $j$th lag).  In addition, the lasso can allow the dependence among exogenous features within a lag to vary. 

\subsection{Tuning Lasso AR-X: Rolling Validation and Nonstationarity Concerns}\label{penaltysection}

In a time-dependent context, penalty parameter selection is not well suited to traditional $n$-fold cross validation.  Instead, \emph{rolling} validation, put forth by \cite{BGR}, selects the optimal penalty parameter as the minimizer of a loss function after iterating forward one observation at a time over a training period.  This procedure divides the data into three periods: one for initialization, one for training, and one for forecast evaluation.  For the purposes of this paper, we choose time indices $T_1=\left \lfloor \frac{T}{3} \right\rfloor,  T_2=\left\lfloor \frac{2T}{3} \right\rfloor$.  In our applications of rolling validation, the period $T_1$ through $T_2-1$ is used for training and $T_2$ through $T$ is a holdout set reserved for evaluation of forecast accuracy.   

Let $\hat{y}_{t+1}^{\lambda}$ be the one-step ahead forecast based on all observations up to time $t$ corresponding to a given penalty parameter $\lambda$ from a predetermined grid of values $[\lambda^{1},\dots,\lambda^{\max}]$.  Rolling  validation then chooses $\hat\lambda$  as the minimizer of one step ahead mean squared forecast error (MSFE) averaged over the training period 
\begin{equation*}
\text{MSFE}(\lambda)=\frac{1}{T_2-T_1}\sum_{t=T_1}^{T_2-1} \left(\hat{y}_{t+1}^{\lambda}-y_{t+1}\right)^2.
\end{equation*}
Other loss functions could be considered, but MSFE is the most natural given our use of a least squares objective function.  One of the major drawbacks of the rolling  validation procedure is its computational burden; it requires solving $T_{2}-T_1$ lasso optimization problems over a grid of $n$ penalty values.  Though the lasso can be solved efficiently via an iterative procedure such as coordinate descent \citep{friedman}, a single pass has complexity O($(p+ks)T$), which can be burdensome if $p,s,k$ and $T$ are large.   

\begin{figure}
\centering
\includegraphics[scale=.66]{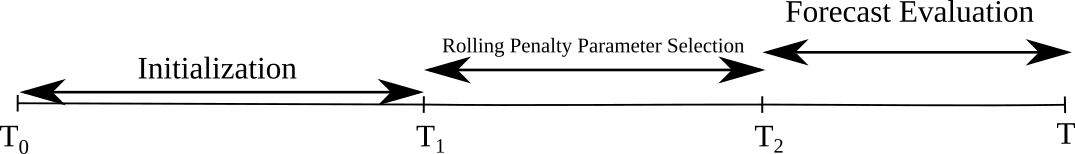}\hspace{1cm}
\includegraphics[scale=.66]{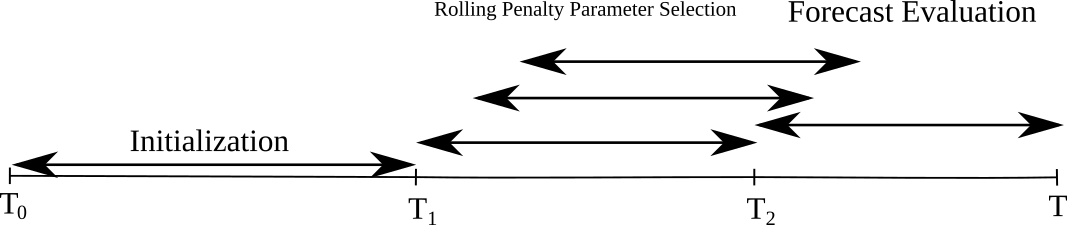}
\caption{Two variations of conventional rolling validation: (Left) Rolling over selection period with fixed penalty parameter over evaluation period and (Right) Rolling over selection period with continuous tuning as new data point is realized (over the original set of candidate $\lambda$).}
\label{fig:figcva1}
\end{figure}  

Two variations of this procedure are illustrated in Figure \ref{fig:figcva1}.  Each approach uses the period from  $T_1$ to $T_2-1$ to obtain the initial penalty parameter.  The approach in the left panel evaluates this fixed penalty parameter over a holdout set and is often used for purposes of relative forecast comparison.  The right panel introduces an approach that is more adaptive to changes in the data, as it rolls the penalty parameter selection window forward one observation and re-evaluate the optimal penalty parameter over the original set of candidate penalty parameters.  


Under many scenarios, selection with a fixed grid of penalty parameters may not be appropriate.  For example, most time series models implicitly assume that the series is \emph{stationary}, which, as defined by \cite{shumway}, requires that the conditional mean of the series is constant across all observations and the covariance between two observations $t_0$ and $t_1$ is a function of their distance ($|t_1-t_0|$).  Such assumptions are very restrictive and eliminate series exhibiting trend-like behavior as well as those with a substantial degree of variability.  

There is a widespread belief that most financial \citep{Fama} and macroeconomic \citep{Litterman1979} time series exhibit nonstationary behavior.  It is common practice in macroeconomic forecasting \citep{stock2002macroeconomic} to transform series to stationarity by differencing.  However, as stated in \cite{Hong} nonstationarity in macroeconomic time series can be driven by structural shocks including policy regime shifts, technological progress, and changes in preferences which cannot be adequately addressed simply by differencing.   

Under nonstationarity, conventional rolling validation is inappropriate, as the optimal degree of regularization may vary at different points in the series.  In the next section, we detail an online approach that can more accurately capture the nonstationary tendencies of these time series.    

\section{Proposed Framework}
\label{sec:sec3}
We start by  describing the recursive lasso algorithm adopted from \cite{Garr} used for online updating of the lasso solution. Utilizing this algorithm, we propose a penalty parameter selection scheme that combines the conventional rolling  approach with an online updating scheme.  For the ease of notation, we re-parameterize the lasso AR-X problem from Section \ref{sec:sec2} into a compact matrix format.  Let $\bm{z}_{i}=[y_{i-1},\ldots,y_{i-p},x_{1,i-1},\ldots,x_{k,i-s}]^{\top}$
represent all lagged observations at time $i\ge T_0$. Let $\PhiB=[\phi_1,\dots,\phi_p,\dots,\theta_{11},\dots,\theta_{ks}]^{\top}$ denote the vector of coefficients.  Then, we can express the lasso AR-X problem at time $t$ as
\begin{equation}\label{newprob}
\PhiB^{(t, \lambda)}=\argmin_{\PhiB\in\mathbb R^{(p+ks)}} \left\{\frac{1}{2} \sum_{i=T_0}^t \left(y_i-\bm{z}_i^{\top}\PhiB\right)^2+\lambda\|\PhiB\|_1\right\}. 
\end{equation}
Let $\bm{y}=[y_{T_0},\dots, y_t]^{\top}$ and $\Z\in\mathbb R^{(t-T_0+1)\times (p+ks)}$ be the matrix with the rows being $\{\bm{z}_{T_0}^{\top}, \ldots, \bm{z}_t^{\top}\}$. Let $A\subset \{1, \ldots, p+ks\}$ be the set of indices for nonzero elements of $\PhiB$, i.e., its \emph{active set}. Let $\PhiB_A\subset \PhiB$ denote the active features. We partition $\Z$ so that the columns of $\Z_A$ correspond to the active set. Let $\bm{v}\in \{-1, 0, 1\}^{(p+ks)}$ be the sign set of $\PhiB$ such that $v_j=\text{sign}(\Phi_{j})$ for all $j$. By standard optimality conditions in a convex problem \citep{boyd2004}, we have $\PhiB^{(t, \lambda)}_A=(\Z_A^{\top}\Z_A)^{-1}(\Z_A^{\top}\bm{y}-\lambda \bm{v}_A)$ as the active set solution to Problem \eqref{newprob} at time $t$.

\subsection{Online Lasso Update with RecLasso}
\label{sec:reclasso}
Online updating is performed throughout the period $[T_2, T)$, extending the Recursive Lasso (RecLasso) algorithm from \cite{Garr} to a time dependent setting. RecLasso is heavily influenced by the Least Angle Regression (LARS) algorithm proposed by \cite{efron}, which computes the entire regularization path by decreasing the penalty parameter as new features of $\PhiB$ enter and leave.  RecLasso greatly reduces computational requirements by keeping track of changes in the set of nonzero elements of $\PhiB$. In practice, when $\PhiB$ is sparse and its nonzero elements do not change much as a new observation is added, RecLasso is much more efficient than coordinate descent.

Now, with the new observation $(y_{t+1}, \bm{z}_{t+1})\in\mathbb R\times\mathbb R^{(p+ks)}$ and corresponding penalty $\lambda_{t+1}$, \cite{Garr} introduce the following augmented problem in aid of the computation of the updated coefficients $\PhiB^{(t+1, \lambda_{t+1})}$:
\begin{equation}\label{augprob}
\PhiB(\gamma, \lambda) = \argmin_{\PhiB\in\mathbb R^{(p+ks)}}\left\{\frac{1}{2}\left\| \begin{pmatrix} \bm{y}\\\gamma y_{t+1} \end{pmatrix}-\begin{pmatrix} \Z\\\gamma \bm{z}_{t+1}^{\top} \end{pmatrix}\PhiB \right\|_2^2 +\lambda \|\PhiB\|_1\right\}.
\end{equation}
With the augmented problem, one can equivalently express $\PhiB^{(t, \lambda_{t})}=\PhiB(0,\lambda_{t})$ and $\PhiB^{(t+1, \lambda_{t+1})}=\PhiB(1,\lambda_{t+1})$. We summarize RecLasso that computes a path from $\PhiB^{(t, \lambda_t)}$ to $\PhiB^{(t+1, \lambda_{t+1})}$ in two steps in Algorithm \ref{RecLassoAlgo}. The efficiency of RecLasso is contributed by its focus on \emph{transition points} which correspond to values of $\gamma\in [0, 1]$ at which the active set  changes (i.e., a new feature is added to the active set or an existing active feature is removed). We refer to Section 3.2 of \cite{Garr} for details of computing transition points.
\begin{algorithm}
\caption{RecLasso Algorithm \citep{Garr}}
\label{RecLassoAlgo}
\begin{algorithmic}[1]
\REQUIRE Compute the path from $\PhiB(0,\lambda_{t})$ to $\PhiB(0,\lambda_{t+1})$ using LARS.
\ENSURE Compute the path from $\PhiB(0,\lambda_{t+1})$ to $\PhiB(1,\lambda_{t+1})$ as the following.
\STATE Let $\bm{v}$ be the sign set of $\PhiB(0,\lambda_{t+1})$. Let $\tilde{\bm{y}}=\begin{pmatrix}\bm{y}\\ y_{t+1}  \end{pmatrix}$ and $\tilde{\Z}=\begin{pmatrix} \Z\\  \bm{z}_{t+1}^{\top}\end{pmatrix}$.
\STATE Initialize the active set $A$ to the indices of the nonzero coefficients of $\PhiB(0,\lambda_{t+1})$. Initialize the transition point $\gamma=0$ and the active set features 
\begin{equation}\label{RecLassoSoln}
\tilde \PhiB_A=(\tilde{\Z}^{\top}_A\tilde{\Z}_A)^{-1}(\tilde{\Z}_A^{\top}\tilde{\bm{y}}-\lambda_{t+1}\bm{v}_A).\vspace{-1.5em}
\end{equation}
\STATE Keep computing the next transition point $\gamma$ until $\gamma>1$. At each transition point, update accordingly the active set $A$, $\bm{v}_A$, $\tilde{\Z}_A$ and $\tilde\PhiB_A$ (by \eqref{RecLassoSoln}).
\STATE Return $\tilde \PhiB_A$ as the values of $\PhiB(1,\lambda_{t+1})$ on its active set.
\end{algorithmic}
\end{algorithm}

Note that since at each transition point, at most one feature can leave or enter the active set, we do not need to explicitly compute a matrix inverse when updating $\tilde{\PhiB}_A$.  We instead utilize the Sherman-Morrison formula \citep{sherman} to perform rank one updates to $(\tilde{\Z}^{\top}_A\tilde{\Z}_A)^{-1}$.

\subsection{Online Penalty Parameter Selection for Lasso AR-X}
\label{sec:secadapt}

As mentioned in Section \ref{penaltysection}, rolling  validation, the conventional penalty parameter selection for regularized time series models, is based on minimizing the MSFE over the training period.  Since the first step of RecLasso algorithm involves varying the penalty parameter as a new observation enters, it is very amenable to an online updating scheme. Our proposed online regularization methods address two drawbacks to rolling validation. First, rolling validation is very computationally intensive. Second, a fixed grid of penalty values cannot accurately account for nonstationarity.  \cite{Garr} describe an approach that iteratively determines the penalty parameter using each point in the training data as a test set. Their online regularization process successively determines the amount of regularization in a data-driven manner at a substantially lower computational overhead than conventional methods. However, in practice we observed that the selected penalty parameter depends heavily on the starting value of this online process, which consequently affects the performance of our estimator.

In order to obtain an adequate starting value for our penalty parameter, we perform rolling validation over the training period $[T_1, T_2)$ on a fixed grid of penalty parameters. We use the parameter $\hat{\lambda}$ selected from rolling validation as a starting value and iteratively update the penalty value over the evaluation period for a new observation and record the out of sample MSFE for observation at $t+1$. The procedure is illustrated in Figure \ref{fig:figcvadapt}.  

\begin{figure}[H]
\centering
\label{fig:figcvadapt}
\includegraphics[scale=.66]{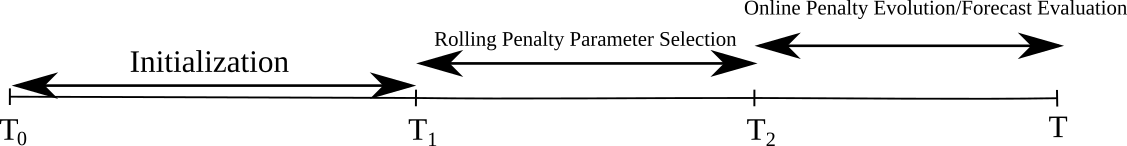}.
\caption{Illustration of the online regularization procedure for lasso AR-X}
\label{fig:figcvadapt}
\end{figure}  

We propose two updating rules for our online updating procedure in Sections \ref{sec:gradientdescent} and \ref{sec:regu-newton}. Let $\hat\lambda_{T_2}$ be the optimal penalty selected from rolling  validation in the training period. For $t\in[T_2,T)$, let $\hat\lambda_t$ be the penalty value at point $t$.  For out-of-sample comparison purposes, we first record the MSFE of observation at $t+1$ with $\hat\lambda_t$.  We then use this observation at $t+1$ as a ``test point'' and update to $\hat\lambda_{t+1}$ in the direction that minimizes the prediction error 
\begin{equation}\label{errorfunc}
err(\lambda):=\left(y_{t+1}-\bm{z}_{t+1}^{\top}\PhiB^{(t, \lambda)}\right)^2
\end{equation}
where $\PhiB^{(t, \lambda)}$ is the solution to Problem \eqref{newprob}. We use gradient descent or the Newton method to determine specifics of the iterative updates, as described in details below.

\subsubsection{Gradient Descent Online Regularization}\label{sec:gradientdescent}
Following the proposal in \cite{Garr}, we consider using gradient descent to update $\lambda$.  Similar to Section \ref{sec:reclasso}, we let  $\PhiB_A$,  $\Z_A$ and $\bm{v}_A$ correspond to the active set $A$ at time $t$. For the next observation $(y_{t+1}, \bm{z}_{t+1})$, we partition $\bm{z}_{t+1}$ so that $\bm{z}_{t+1, A}$ corresponds to the active set. The error function for the observation at $t+1$ is
\begin{equation}
err(\lambda) =(\bm{z}_{t+1}^{\top}\PhiB^{(t, \lambda)}-y_{t+1})^2
=(\bm{z}_{t+1,A}^{\top}\PhiB_A^{(t, \lambda)}-y_{t+1})^2
\end{equation}
where $\PhiB_A^{(t, \lambda)}=(\Z_A^{\top}\Z_A)^{-1}(\Z_A^{\top}\bm{y}-\lambda \bm{v}_A)$. The subdifferential $\nabla err(\lambda)$ with respect to $\log \lambda$ is 
\begin{align*}
\frac{\partial err(\lambda)}{\partial \log \lambda}&= 2\left(\bm{z}_{t+1,A}^{\top}\PhiB_A^{(t, \lambda)}-y_{t+1}\right)\cdot \left( -\bm{z}_{t+1,A}^{\top}(\Z_A^{\top}\Z_A)^{-1} { \lambda} \bm{v}_A\right)\\
&=-2\lambda\bm{v}_A\bm{z}_{t+1,A}^{\top}(\Z_A^{\top}\Z_A)^{-1}\left(\bm{z}_{t+1,A}^{\top}\PhiB_A^{(t, \lambda)}-y_{t+1}\right).
\end{align*}
Suppose we use gradient descent to minimize $err(\lambda)$ on a $\log$-scale of $\lambda$ with the current parameter being $\hat\lambda_t$. Following its updating rule we have $\log(\hat\lambda_{t+1}) \leftarrow\log(\hat\lambda_t)-\eta\frac{\partial err(\hat\lambda_t)}{\log\hat\lambda}$, i.e., 
\begin{equation*}
 \hat\lambda_{t+1} \leftarrow\hat\lambda_t\times\exp\left\{2 \eta \hat\lambda_t\bm{v}_A\bm{z}_{t+1,A}^{\top}(\Z_A^{\top}\Z_A)^{-1}\left(\bm{z}_{t+1,A}^{\top}\PhiB_A^{(t, \hat\lambda_t)}-y_{t+1}\right)\right\}
\end{equation*}
where $\eta$, the learning rate controlling the stepsize, is small; in our simulations and data application in Section \ref{sec:sec4}, we set it to 0.1. Finally, we perform the update in the $\log$ domain to ensure that $\hat\lambda$ is always positive.

\subsubsection{Newton Online Regularization}\label{sec:regu-newton}
We additionally propose an updating rule that uses the Newton method to minimize $err(\lambda)$ on the $\log$-scale of $\lambda$. The Newton method takes curvature into account: its stepsize is inversely related to the ``steepness,'' which should make its update more adaptive than using a fixed step size.  The Newton method requires the computation of the Hessian matrix of $err(\lambda)$:
\begin{equation*}
H_{err}(\lambda)=\frac{\partial^2 err(\lambda)}{\partial (\log\lambda)^2}
=-2\lambda \bm{v}_A \bm{z}_{t+1,1}^{\top}(\Z_A^{\top}\Z_A)^{-1}\left(\bm{z}_{t+1,A}^{\top}\boldsymbol\Sigma(\lambda)-y_{t+1}  \right)
\end{equation*}
where $\boldsymbol\Sigma(\lambda)=(\Z_A^{\top}\Z_A)^{-1}(\Z_A^{\top}\bm{y}-2\lambda \bm{v}_A)$. Following Newton's updating rule we have $\log(\hat\lambda_{t+1}) \leftarrow \log(\hat\lambda_t) -\frac{\nabla err(\hat\lambda_t)}{H_{err}(\hat\lambda_t)}$, i.e., 
\begin{equation*}
\hat\lambda_{t+1} \leftarrow \hat \lambda_t \times\exp\left\{ \frac{\bm{z}_{t+1,1}^{\top}\PhiB_A^{(t, \hat\lambda_t)}-y_{t+1}}{\bm{z}_{t+1,1}^{\top}\boldsymbol \Sigma(\hat\lambda_t)-y_{t+1}}  \right\}.
\end{equation*}
We include details for deriving Newton's updating rule in Appendix \ref{appen:newton}.

\section{Experiment Results}
\label{sec:sec4}
In this section, we compare the forecasting and computational performance of our online procedures with conventional IC methods as well as the standard lasso AR-X (with the penalty parameter selected by rolling  validation).  We present both a simulation example and a macroeconomic data application.  All of our analysis was performed in {\tt R} \citep{R}.  Our RecLasso implementation was heavily influenced by the Python code made publicly available by \cite{Garr} at \url{https://github.com/pierreg/reclasso}; the Lasso AR-X and other benchmarks were fit using BigVAR \citep{BigVAR}.      

\subsection{Simulation}
\begin{figure}[ht]
\centering
\label{fig:figsp}
\includegraphics[scale=.75]{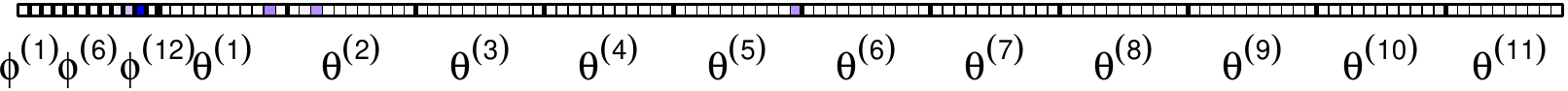}. 
\caption{Sparsity pattern of true coefficient structure (active coefficients shaded) used in simulation.  Darker coefficients are larger in magnitude.}
\label{fig:sparsity}
\end{figure}

We consider simulating from an AR-X model with $k=10$, $p=12$ and $s=12$ (143 potential features) with series length $T=250$.  The simulation structure was generated to create nonzero coefficients at random while ensuring that the resulting series is stationary.  The sparsity pattern of this simulation structure is depicted in Figure \ref{fig:sparsity}.    

We compare the forecasting performance of our online regularization approaches with the two variants of the conventional lasso AR-X with the penalty parameter selected according to rolling validation described in Section \ref{penaltysection}.  We also compare against several conventional approaches: the unconditional sample mean, the random walk, and two information criterion based measures. The unconditional sample mean forecasts $\hat{y}_{t+1}$ using the mean of all observations up to time t.  The random walk simply forecasts $\hat{y}_{t+1}=y_t$.  We additionally consider fitting an AR-X by least squares with the lag order selected according to minimization of AIC and BIC (note that maximal lag orders are restricted to ensure that $ks+p<T$).  Table \ref{tab:tab2} reports the resulting MSFE averaged over 100 simulations.  
     
{\small
       \begin{table*}
    \centering    
    \small
 \caption{\label{tab:tab2}Out of sample MSFE of one-step ahead forecasts averaged over 100 simulations (relative to fixed grid lasso AR-X)}
   \begin{tabular}{ l | c c }
\hline
Model & Relative MSFE & Standard Error \\
     \hline
Lasso AR-X - Static Eval Period   & 1.0000 & 0.00241\\
Lasso AR-X - Rolling Window Eval Period    & 0.9921 & 0.00234\\     
Online Lasso AR-X - Gradient Descent     & {\bf 0.9813} & 0.00225\\
Online Lasso AR-X - Newton      & 0.9845 & 0.00226\\
Sample Mean & 2.8367  &0.00856 \\
Random Walk & 4.4431  &0.00856 \\
AR-X with lag order selected by AIC & 5.1585& 0.02049\\
AR-X with lag order selected by BIC & 5.4009 & 0.02049\\
\hline
  \end{tabular}
 \end{table*}  
}

As expected, the lasso procedures substantially outperform the IC based approaches as well as the sample mean and random walk.  Both the gradient descent and the Newton based online lasso achieve superior performance to the rolling window lasso which subsequently outperforms the lasso with a static evaluation period.

\begin{figure}[ht]
\centering
\includegraphics[width = 0.5\linewidth]{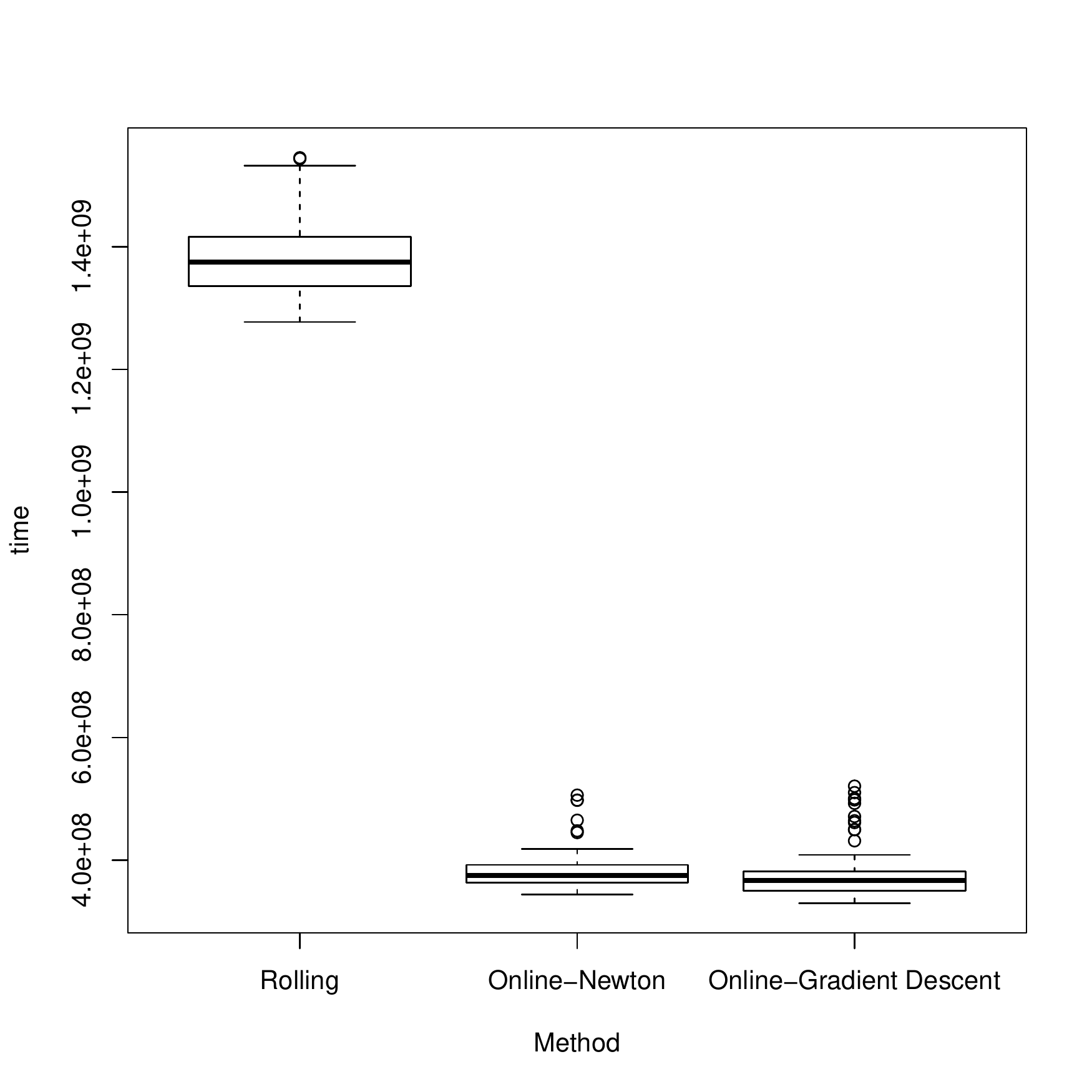}
\caption{Boxplots comparing the computational times of rolling validation and both online penalty parameter selection procedures. }
\label{fig:fig3}
\end{figure}

In terms of computational performance, we compare our online lasso procedures against conventional rolling validation over the training period with $T_2-T_1=76$. Using the {\tt microbenchmark} package in R, we compare the average computational time of one full iteration of rolling validation against both versions of the online lasso over the same period with the same starting penalty parameter. The results are recorded in Table \ref{tab:tab1} and Figure \ref{fig:fig3}. Despite the rolling validation implementation utilizing a warm start and a highly optimized {\tt C++} implementation, both online lasso procedures are substantially faster.

\begin{table*}
\centering
\small
\caption{\label{tab:tab1} Distribution of computational times (in milliseconds) of 100 iterations of rolling validation compared to both online penalty parameter selection procedures. }
\begin{tabular}{l|rrrrrr}
 \hline
 Lasso AR-X Model & Min & Lower Quartile & Mean & Median & Upper Quartile & Max  \\ 
  \hline
 Conventional Rolling &   1266.31 & 1277.94 & 1306.63 & 1288.18 & 1309.65 & 1527.03 \\ 
 Online - Newton &  345.45 & 355.25 & 367.76 & 359.25 & 367.23 & 490.37  \\ 
Online - Gradient Descent & 334.89 & 343.83 & 359.44 & 348.95 & 355.62 & 472.98 \\
   \hline
\end{tabular}
\end{table*}

\subsection{Data Application}
We additionally evaluate our procedures on a set of US macroeconomic indicators procured from the databases FRED-QD \citep{fredqd} and FRED-MD \citep{fredmd}.  All monthly indicators were aggregated to quarterly and all series were transformed to stationarity using the provided transformation codes.  These series range from Quarter 2 of 1960 to Quarter 4 of 2019.  We forecast three series, US Industrial Production (GDPC1), a measure of economic activity, the Consumer Price Index (CPIAUCSL), a measure of inflation, and the Federal Funds Rate (FEDFUNDS), a measure of monetary policy.  Procedures that can accurately forecast these three series are of substantial interest to both macroeconomists and policymakers.  These series are plotted in the left panel of Figure \ref{fig:figplots4}.  Note that despite the transformations, evidence of nonstationarity remains as the variability in the late 1970s and early 1980s is abnormally high compared to other periods.

\begin{figure}
\centering
\label{fig:figplots4}
\includegraphics[width=.49\linewidth]{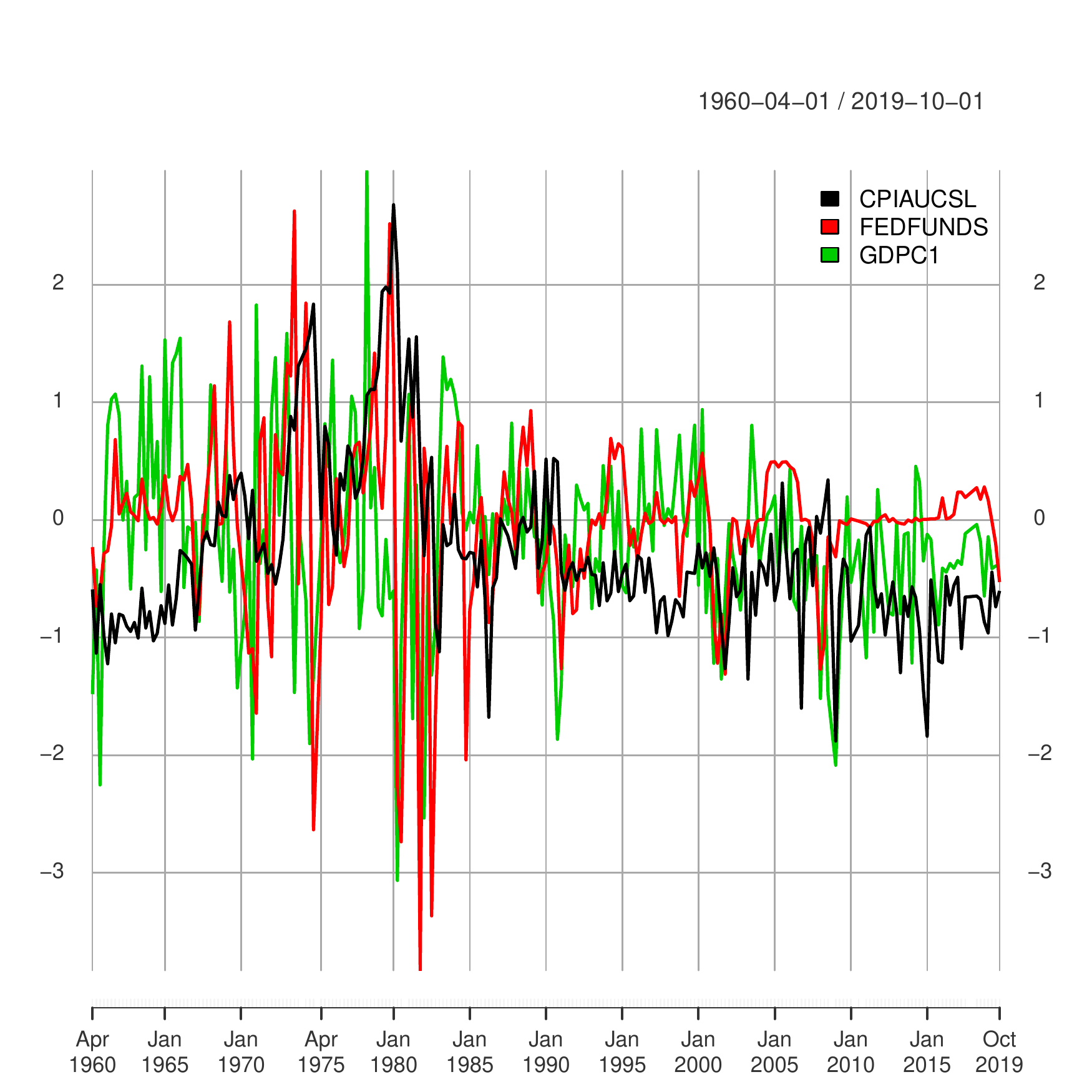}
\includegraphics[width=.49\linewidth]{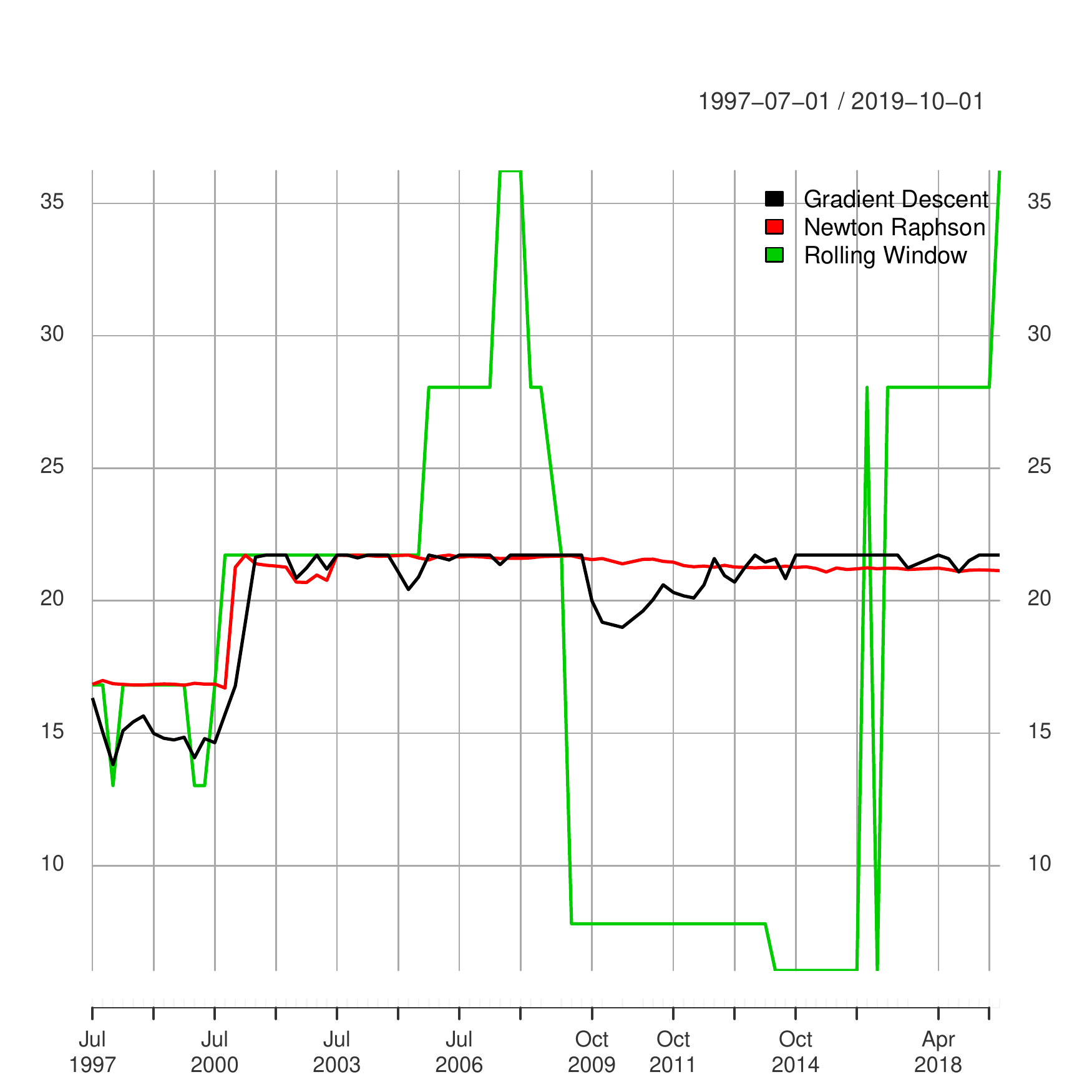}
\caption{(Left) Plots of CPIAUCSL, GDPC1 and FEDFUNDS time series transformed to stationary and normalized to have zero mean and unit variance. (Right) Comparison of the evolution of lasso penalty parameters for the Industrial Production time series. }
 \label{fig:figplots4}
\end{figure}

We consider forecasting these three series using their own past lags as well as the past lags of 86 additional related macroeconomic series as exogenous variables (Groups 2 and 7 from FRED-MD and Group 1 from FRED-QD).  We set $p=s=12$, resulting in 1068 potential features.  Quarter 2 of 1988 to Quarter 2 of 1997 is used for penalty parameter selection, while Quarter 3 of 1997 to Quarter 4 of 2019 is used for forecast evaluation.  Our results are summarized in Table \ref{tab:tab3}.
{\small
       \begin{table*}[th]
    \centering  
    \small  
 \caption{Out of sample MSFE of one-step ahead forecasts of three macroeconomic indicators (relative to fixed grid lasso AR-X)}
\label{tab:tab3}
   \begin{tabular}{ l | c | c| c}
\hline
Model/Series & \begin{tabular}{c}FEDFUNDS \\ Relative MSFE\end{tabular} & \begin{tabular}{c}CPIAUCSL \\ Relative MSFE\end{tabular} & \begin{tabular}{c}GDPC1 \\ Relative MSFE\end{tabular}\\
    \hline
Lasso AR-X - Static Eval Period  & 1.0000 & 1.0000 & 1.0000 \\
Lasso AR-X - Rolling Window Eval Period & 1.0140 & 1.0735 & 1.0745 \\
Online Lasso AR-X - Gradient Descent     & {\bf 0.8840} & {\bf 0.9678} & 0.9390 \\
Online Lasso AR-X - Newton    & 0.9477 & 0.9945 & {\bf 0.9298} \\
Sample Mean & 1.0703 & 2.1341 & 2.6081  \\
AR-X with lag selected by BIC &3.1522 &1.4889 &1.3010\\
AR-X with lag selected by AIC &2.7406 &1.7106 &1.3052\\
\hline
  \end{tabular}
 \end{table*}  
}

We find that the online lasso methods outperform both the fixed grid and rolling window lasso AR-X for each series which subsequently outperform every other method.  In addition, the lasso AR-X that uses the rolling window never outperforms the lasso AR-X that uses the fixed grid.  

The right panel of Figure \ref{fig:figplots4} compares the evolution of the penalty parameter across the gradient descent, Newton, and rolling approaches.  Both the gradient descent and Newton follow a similar evolution with the Newton slightly smoother, perhaps due to the relatively high learning rate used for gradient descent.  Since the rolling approach is restricted to the original penalty grid it makes fewer changes, but they are larger in magnitude.

\section{Conclusion}
\label{sec:sec5}
The results from our simulation and empirical applications are encouraging as we see that both the Newton and gradient descent variants of the online lasso AR-X substantially outperform the conventional fixed grid and rolling window lasso AR-X.  This provides evidence that our approaches can accurately forecast with varying degrees of temporal dependence.  

Our work still has considerable room for extensions.  Future applications could consider a formalization of the penalty parameter selection procedure as well as stricter rules regarding update steps.  In addition, as an alternative to the lasso, certain applications may be more amenable to the elastic net \citep{elastic} or structured penalties such as the group lasso \citep{Yuan}.

\appendix
\section*{Appendices}
\addcontentsline{toc}{section}{Appendices}
\renewcommand{\thesubsection}{\Alph{subsection}}

\subsection{Derivation of Newton Online Regularization}\label{appen:newton}
Here we provide details on deriving the online regularization procedure using the Newton method. The error function for the observation at $t+1$ is 
\begin{align*}
err(\mu) &=(\bm{z}_{t+1, A}^{\top}\PhiB^{(t, \lambda)}_A-y_{t+1})^2 \\
&=\left( \bm{z}_{t+1,A}^{\top}(\bm{Z}_A^{\top}\bm{Z}_A)^{-1}(\bm{Z}_A^{\top}\bm{y}-\exp(\log\lambda)  \bm{v}_A)-y_{t+1}  \right)^2.
\end{align*}
where $\PhiB_A^{(t, \lambda)}=(\Z_A^{\top}\Z_A)^{-1}(\Z_A^{\top}\bm{y}-\lambda \bm{v}_A)$. The subdifferential of $err(\mu)$ with respect to $\log\lambda$ is 
\begin{align}
\nabla err(\lambda) &=\frac{\partial err(\lambda)}{\partial \log\lambda} \nonumber\\
&= 2\left(\bm{z}_{t+1,A}^{\top}\PhiB_A^{(t, \lambda)}-y_{t+1}\right)\cdot \left( -\bm{z}_{t+1,A}^{\top}(\Z_A^{\top}\Z_A)^{-1} { \lambda} \bm{v}_A\right)\nonumber\\
&=-2\lambda\bm{v}_A\bm{z}_{t+1,A}^{\top}(\Z_A^{\top}\Z_A)^{-1}\left(\bm{z}_{t+1,A}^{\top}\PhiB_A^{(t, \lambda)}-y_{t+1}\right).
\end{align}
The Hessian of $err(\lambda)$ with respect to $\log\lambda$ is 
\begin{align}
H_{err}(\lambda)&=\frac{\partial^2 err(\lambda)}{\partial (\log\lambda)^2}\nonumber\\
&= -2\lambda \bm{v}_A \bm{z}_{t+1,A}^{\top}(\bm{Z}_A^{\top}\bm{Z}_A)^{-1}\left( \bm{z}_{t+1,A}^{\top}(\bm{Z}_A^{\top}\bm{Z}_A)^{-1}(\bm{Z}_A^{\top}\bm{y}-\lambda\bm{v}_A)-y_{t+1} - \bm{z}_{t+1,A}^{\top}(\bm{Z}_A^{\top}\bm{Z}_A)^{-1}\lambda\bm{v}_A\right)\nonumber\\
&= -2\lambda \bm{v}_A \bm{z}_{t+1,A}^{\top}(\bm{Z}_A^{\top}\bm{Z}_A)^{-1}\left( \bm{z}_{t+1,A}^{\top}(\bm{Z}_A^{\top}\bm{Z}_A)^{-1}(\bm{Z}_A^{\top}\bm{y}-2\lambda\bm{v}_A)-y_{t+1}\right)\nonumber\\
&= -2\lambda \bm{v}_A \bm{z}_{t+1,A}^{\top}(\bm{Z}_A^{\top}\bm{Z}_A)^{-1}\left( \bm{z}_{t+1,A}^{\top}\boldsymbol\Sigma(\lambda)-y_{t+1}\right)
\end{align}
where $\boldsymbol\Sigma(\lambda)=(\Z_A^{\top}\Z_A)^{-1}(\Z_A^{\top}\bm{y}-2\lambda \bm{v}_A)$. Using the Newton method to minimize $err(\lambda)$ at the $\log$-scale of $\lambda$, we have $\log(\hat\lambda_{t+1}) \leftarrow \log(\hat\lambda_t) -\frac{\nabla err(\hat\lambda_t)}{H_{err}(\hat\lambda_t)}$, i.e., 
\begin{equation*}
\hat\lambda_{t+1} \leftarrow \hat \lambda_t \times\exp\left\{ \frac{\bm{z}_{t+1,1}^{\top}\PhiB_A^{(t, \hat\lambda_t)}-y_{t+1}}{\bm{z}_{t+1,1}^{\top}\boldsymbol \Sigma(\hat\lambda_t)-y_{t+1}}  \right\}.
\end{equation*}

\bibliography{harbib}
\bibliographystyle{apalike}
\end{document}